\documentclass{nime-alternate}

\begin{document}
%
\conferenceinfo{NIME'17,}{May 15-19, 2017, Aalborg University Copenhagen, Denmark.}

\title{Cross-Modal Terrains: Navigating Sonic Space through Haptic Feedback}

%
%
%
%
%

\numberofauthors{3} 
%
\author{
%
%
\alignauthor
Gabriella Isaac\\
    \affaddr{Arts, Media + Engineering} \\
    \affaddr{Arizona State University} \\
    \affaddr{Tempe, AZ 85287, USA} \\
    \email{Gabriella.Isaac@asu.edu}
\alignauthor
Lauren Hayes\\
    \affaddr{Arts, Media + Engineering} \\
    \affaddr{Arizona State University} \\
    \affaddr{Tempe, AZ 85287, USA} \\
    \email{lauren.s.hayes@asu.edu}
\alignauthor Todd Ingalls\\
     \affaddr{Arts, Media + Engineering} \\
    \affaddr{Arizona State University} \\
    \affaddr{Tempe, AZ 85287, USA} \\
    \email{Todd.Ingalls@asu.edu}
   }


\maketitle
\begin{abstract}
This paper explores the idea of using virtual textural terrains as a means of generating haptic profiles for force-feedback controllers. This approach breaks from the para-digm established within audio-haptic research over the last few decades where physical models within virtual environments are designed to transduce gesture into sonic output.  We outline a method for generating multimodal terrains using basis functions, which are rendered into monochromatic visual representations for inspection. This visual terrain is traversed using a haptic controller, the NovInt Falcon, which in turn receives force information based on the grayscale value of its location in this virtual space. As the image is traversed by a performer the levels of resistance vary, and the image is realized as a physical terrain. We discuss the potential of this approach to afford engaging musical experiences for both the performer and the audience as iterated through numerous performances.
\end{abstract}


\acmclassification1998{
H.5.5 [Information Interfaces and Presentation] Sound and Music Computing, H.5.2 [Information Interfaces and Presentation] User Interfaces---Haptic I/O.

}

\section{Introduction}
The introduction of digital musical instruments (DMIs) has removed the need for the existence of a physically resonating body in order to create music, leaving the practice of sound-making often decoupled from the resulting sound. The inclination towards smooth and seamless interaction in the creation of new DMIs has led to the development of musical instruments and interfaces for which no significant transfer of energy is required to play them. Other than structural boundaries, such systems usually lack any form of physical resistance, whereas the production of sounds through traditional instruments happens precisely at the meeting of the performer's body with the instrument's resistance: ``When the intentions of a musician meet with a body that resists them, friction between the two bodies causes sound to emerge" \cite[p1]{parker15joys}. Haptic controllers offer the ability to engage with digital music in a tangible way. Previous work with force-feedback has derived haptic profiles from physical models of acoustic instruments and real-world interactions (see \cite{berdahl2009hsp} for a discussion of this). Relying on these sources becomes an issue when working with digital signal processes that have no real-world correspondence. By exploring other methods of generating force-feedback profiles, purely digital information can become physically realized in a genuine and engaging manner.

\begin{figure}[htbp]
	\centering	\includegraphics[width=1\columnwidth]{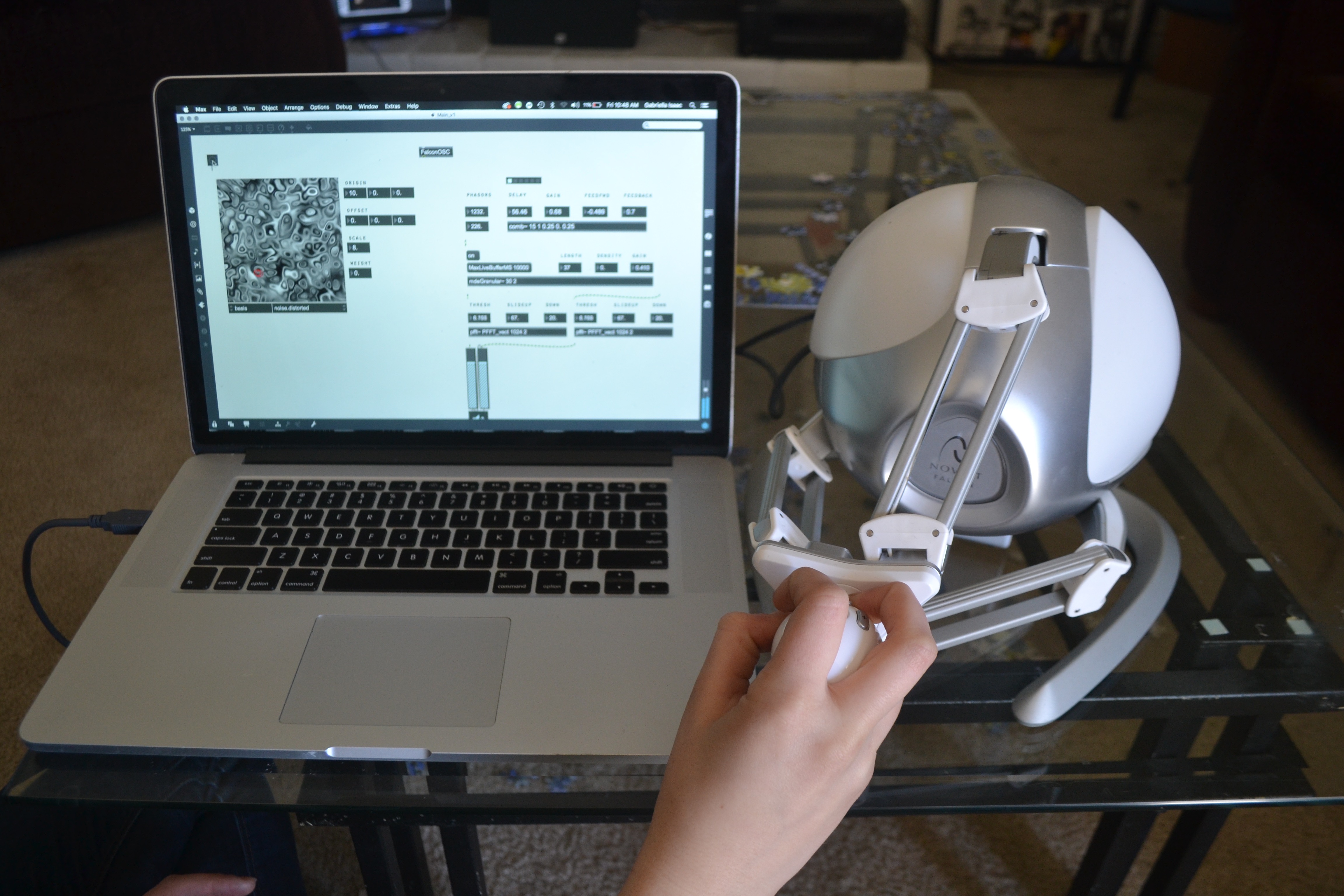}
	\caption{Using basis functions to generate haptic terrains for the NovInt Falcon.}
	\label{fig:gabby}
\end{figure}

\section{Background}
Dynamic relationships occur and are ongoing between the performer, the instrument, and the sounds produced when playing musical instruments. These exchanges depend upon the sensory feedback provided by the instrument in the forms of auditory, visual, and haptic feedback \cite{overholt2011advancements}. Because digital interfaces based around an ergonomic HCI model are generally designed to eliminate friction altogether, the tactile experience of creating a sound is reduced. 
Even though digital interfaces are material tools, the feeling of pressing a button or moving a slider does not provide the performer with much physical resistance, whereas the engagement required to play an acoustic instrument provides musicians with a wider range of haptic feedback involving both cutaneous and proprioceptive information, as well as information about the quality of an occurring sound \cite{overholt2011advancements}. This issue is recognized in Claude Cadoz's work regarding his concept of ergoticity as the physical exchange of energy between performer, instrument, and environment \cite{cadoz1988instrumental}. A possible solution to these issues is the use of haptic controllers. As has been previously noted, ``we are no longer dealing with the physical vibrations of strings, tubes and solid bodies as the sound source, but rather with the impalpable numerical streams of digital signal processing" \cite[p1]{hayes2011vibrotactile}. In physically realizing the immaterial, the design of the force profile is crucial because it determines the overall characteristics of the instrument.

\subsection{Instrumental Modeling}
Previous work on the design of haptic profiles has been focused around modeling, enhancing, or augmenting pre-existing musical interactions. Leonard and Cadoz's physics-based system GENESIS-RT allows users to design virtual musical instruments and engage with them in a way that resembled interaction with traditional instruments \cite{leonard2015physical}. Ber-dahl's work with the physically intuitive haptic drumstick builds off and extends real-world physics, allowing a drummer to play a drum roll either in the usual manner with two drumsticks, or single-handedly with the aid of force-feedback. It is the direct engagement with the synthetic haptic feedback that allows the percussionist to perform in this way yet the ``physics of the performer's basic interaction with the instrument remain similar," rendering the instrument ``physically intuitive" \cite[p363]{berdahl2007physically}.

Further recent developments include actuated instruments that ``produce sound via vibrating element(s) that are co-manipulated by humans and electromechanical systems" \cite[p155]{overholt2011advancements}. Physical instruments such as the Feedback Resonance Guitar, the Electromagnetically Prepared Piano, the Overtone Fiddle, and Teleoperation with Robothands have been extended through virtual qualities while conforming to real-world physics in their structural design. 

\subsection{Sound Sculpting}
Hayes' \textit{Running Backwards, Uphill} was a performance for a piano trio that explored the relationship between ``touch, gesture and timbre by examining the sonic qualities of the acoustic instruments," as well as the electronics. Musicians utilized the absolute extremes of their instruments and were ``directed to lurch and fall off the keys; or, create the most delicate airy bowed sounds" \cite[p402]{hayes2012performing}. Unlike acoustic instruments, which require physical force and highly skilled action to reach the edges of their sonic potential, the extremes of most digital controllers can be reached almost instantaneously due to their lack of resistance \cite[p1]{parker15joys}. With the Falcon's resistance, Hayes was able to use various force-feedback profiles to both assist her in reaching desired effects, such as short, staccato-like samples of a recording and longer segments that flowed together, and fight against her to make certain sounds more difficult to reach. Throughout the performance, she notes that the haptic feedback gave her a ``feeling of moving through the sound" and the means to shape the sound accordingly \cite[p403]{hayes2012performing}. Her use of haptics allowed an extensive range of gestures and levels of engagement to occur in this performance. 

\subsection{Cross-Modal Approaches}
Others have noticed that it is useful to transfer auditory and visual perceptual information to the tactile realm. The Haptic Wave is an interface that allows ``cross-modal mapping of digital to audio to the haptic domain" \cite[p2150]{tanaka2016haptic} and provides a platform for visually impaired users to navigate and edit digital audio. Instead of translating existing representations of digital audio, such as visual waveforms, the initial iterations of this project used the NovInt Falcon to ``directly access qualities of the sound" \cite[p2153]{tanaka2016haptic}. By mapping the horizontal axis of the Falcon to the length of a sound file and the amount of resistance to the amplitude---higher amplitude increased resistance---users were able to scan through the soundfile from left to right and feel the amplitude of the soundfile at every point in time.

Other work has explored the use of visual information within the realm of digital signal processing. James maps 2D visual textures to the distribution of ``timbre spatialization in the frequency domain" \cite[p128]{james2016multi-point}. Filatriau and Arfib's work presents three different tactics for linking ``visual and sonic textures using similar synthesis processes" \cite[p31]{filatriau2006using}. The first two methods use a visual texture as the material basis for sound generation.  Their approach titled ``image from a sound" considers the texture as a sonogram and uses this image to ``derive from a static image an evolving sound" \cite[p32]{filatriau2006using}. This process involves scanning over the static image to generate an evolving sound, using the horizontal axis as the time value and the vertical value at that time as the sinusoid to reproduce. The ``pixel image sonification technique" employs an image that is constantly changing and constantly producing a sound. This could be exemplified by movement in front of a camera or by moving through a static image at a small scaled viewpoint, viewing the visual texture as composed of small sections, kernels, or pixels. The final process of ``equivalent processes" expresses the texture-generating algorithm in visual and sonic mediums instead of using the image as the starting point. Both the visual and sonic processes are expressions of the texture-generating algorithm.

In our work with the Falcon, we have utilized both the ``pixel image sonification" and the ``equivalent processes" technique by using a texture-generating algorithm to create a visual expression and move through this image on a pixel-to-pixel basis. Working in this direction avoids issues of physical modeling because it is not concerned with using images or interactions as true or exact models as a means to expressively sculpt sound.

\section{Methodology}
In this section we describe an iterative approach to the development of a terrain-based haptic instrument. We outline both the technical considerations, along with autoethnographic accounts of its role in numerous performances by Gabriella Isaac over a period of several months.

\subsection{Case Study: Constant Resistance Model}
The first iteration of the instrument involved applying a constant force profile to a single dimension of the NovInt Falcon. This would exert a constant physical force back on the performer's hand in one dimension. In turn, the same z-index of the Falcon's position was mapped to the the start point of a grain's position within a sound file and, by pushing and pulling the ball-grip of the controller, the performer was able to physically scrub through a buffer containing that file while granulating the sound. The force-feedback of the controller remained at a static level throughout the piece and provided a constant background resistance, allowing the performer to fine tune their movement and focus in on points of interest by stabilizing their actions.

In addition, parameters of other DSP processes were map-ped to the x-y position of the Falcon's ball-grip, including the x-y coordinates of an amplitude envelope---which could be drawn in real-time---and the selection and speed of playback of a corpus of samples. Furthermore, the sound output was recorded into a looping buffer and fed to a granulator. Using a button on the ball grip, haptics could be turned on and off: when off, the sound would be recorded into the buffer; when on, recording stopped and granulation started, allowing the performer to hone in on interesting parts of the recorded buffer, and move back through time.
 
This iteration was used in two solo performances and one collaborative performance. In the former, the mappings did not change over time, and therefore the extremes of the piece were reached after about five minutes. Isaac felt that the collaborative performance was more successful as the presence of two performers allowed the sounds produced to occur over a greater period of time and at different intervals. In this context, the limitations of the instrument could be explored further. Additionally, audience feedback from both scenarios suggested that although the audience members could see the performer actively struggling with the Falcon, they did not understand what was actually happening and how the sounds were being produced. With the controller facing forward and the knob pointed towards the performer, the audience could not see the full range of motion exerted during the performance because the body of the controller blocked the view of the performer's gestures.

This first model was a helpful starting point due to the simplicity of the audio-haptic mappings involved. It resulted in a physically-demanding controller, which invited further exploration into the types of performer-instrument relationships it might afford. This approach was also useful in drawing out the sonic characteristics of the specific granular patch that had been implemented.

Isaac found that without the additional resistance, potentially interesting sonic details were more likely to be skimmed over and lost due to the ease of movement. The physical difficulty of moving the controller's ball grip with the added resistance initially provided a more interesting mode of engagement between the performer and the digital sonic material. However, the fixed amount of resistance became too predictable in practice. This suggested that implementing a wider and more dynamic range of resistances should be explored next, along with an investigation into how the relationships between sound and physical resistance could be developed further.

\subsection{Case Study: Haptic Terrain (Version 1)}
\begin{figure}[htbp]
	\centering	\includegraphics[width=1\columnwidth]{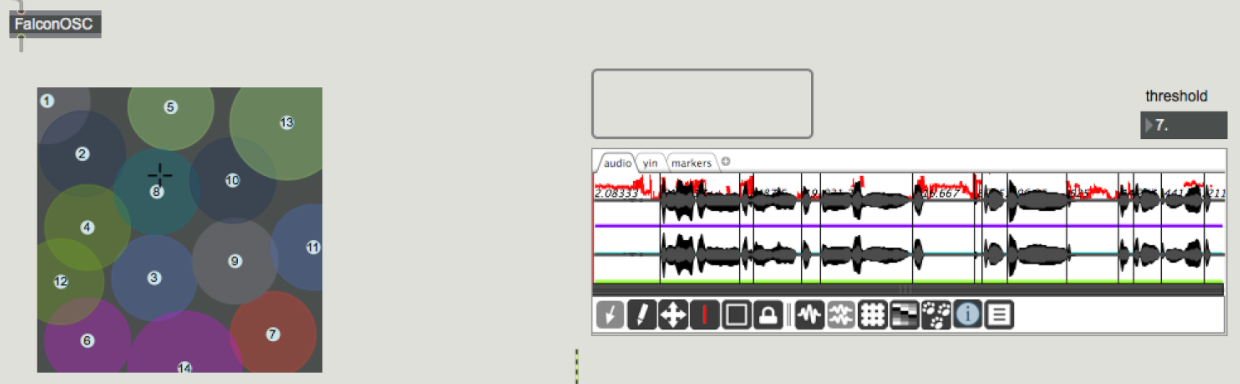}
	\caption{Haptic Terrain Version 1 using nodes and audio segmentation.}
	\label{fig:nodes}
\end{figure}

The second case study builds upon the constant resistance model for using the NovInt Falcon as a controller for performing granular synthesis, and implements an early version of a virtual physical terrain. In this version, rather than working with a constant force in a single dimension, multiple zones of resistance are constructed and placed in arbitrary locations within a virtual 2D space.  Max/MSP's \texttt{nodes} object is a visual panel of circular nodes that can be resized and placed in various locations. A cursor can then be moved through the panel and the object will output whether or not the cursor is within a node's region, as well as its distance from the center of each node.  By mapping the Falcon's x and y position to the coordinates of the cursor, the performer is able to control the cursor's location within a collection of nodes, and traverse the terrain. Using IRCAM's \texttt{MuBu} for Max, a sound file is segmented by an onset threshold and each segment is associated with a specific node (see Figure \ref{fig:nodes}). When the coordinates of the Falcon's ball-grip position are over a node, the specific segment is triggered and sent through a granular synthesizer for further processing. Although the nodes would sometimes overlap due to their varying radii and placement within 2D space, the sound segment selected for playback would always be the segment whose center was closest to the cursor. The output of the node was scaled to the highest resistance value of the Falcon and sent to the force parameter of the z-index---the plane associated with a forward and backward motion for the performer. In this configuration, the performer would receive a low amount of force-feedback towards the outskirts of the node's region, and high amount of feedback towards the center, thus producing the feeling of moving over a bump. 

Because each node was circular, any position at or around the circumference of a node would receive the same amount of feedback. The resistance sent to the Falcon was also scaled and mapped to the density of the grains being played. This meant that if the performer wanted to create a more intense and abrasive sound, they would have to hold the Falcon in the most physically resistant spot. In performance Isaac noted that while this was difficult, it still remained a stable and predictable terrain. While the bumpy resistance created through the nodes improved on the constant resistance model, Isaac felt that this still became predictable when performing because the textures were essentially the same, differing only in scale. This iteration lacked different types of textures and therefore variation in the qualities of the force-feedback. It offered a completely different sensation than the original patch, yet it remained a pre-determined and somewhat stable terrain. Despite this, the nodes could be used in future iterations to access different regions of internal processes within the patch. This iteration could also be used as potential material for an additional visual element for the audience.

\subsection{Case Study: Haptic Terrain (Version 2)}
\begin{figure}[htbp]
	\centering	\includegraphics[width=1\columnwidth]{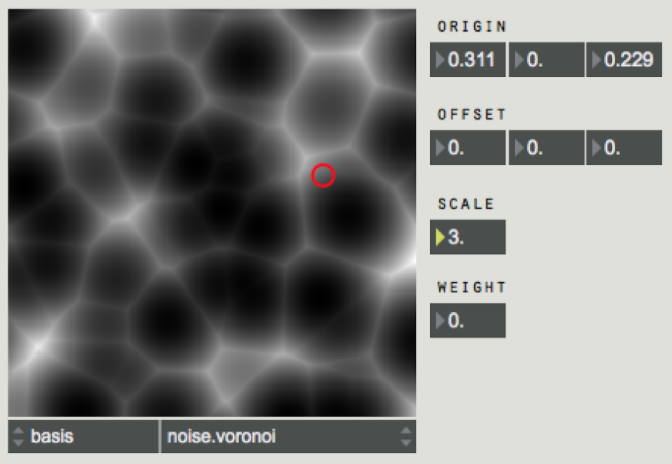}
	\caption{A haptic terrain  generated using Voronoi basis functions.}
	\label{fig:voronoi}
\end{figure}

The second iteration of this idea uses the Max/MSP object \texttt{jit.bfg} to create terrains based on different noise textures. The \texttt{jit.bfg} object ``evaluates and exposes a library of procedural basis functions"\footnote{See the \texttt{jit.bfg} reference page in the Max/MSP documentation.} and by changing the noise basis, different textures can be generated. The  basis function that generated Voronoi cells was favoured during initial tests because, on a close up zoom, it creates crater-like textures with ridges and dips (see Figure \ref{fig:voronoi}). On a higher scale (zoom out), the cells become tiny, bumpy surfaces with defined edges.

The \texttt{jit.bfg} object outputs a grid of texture that is rendered from a range of values from 0 to 1. This range was first normalized, and then mapped onto the maximum and minimum force-feedback values for the NovInt Falcon. Again, the x and y coordinates received from the location of the Falcon's ball-grip were used to scroll through the cells of the 2D matrix generated by \texttt{jit.bfg} and return the greyscale value of that cell. The scaled value was mapped to the force-feedback input of the Falcon. By mapping the highest force of the Falcon to the highest value---the lightest areas of the texture map---and the lowest values to the lowest force of the Falcon---the darkest areas on the texture map---the computationally generated virtual textures were physically realized through haptic interface (see Figure \ref{fig:gabby}).
 
The sonic result of this patch follows the texture maps generated. The source of the sound is generated by two phasors with controllable frequencies that are run through a comb filter. The x and y coordinates change the delay and feedforward of the comb filter. As such, the sound changes drastically depending on the controller position, yet maintains some recognizable characteristics. These sounds are then processed further using granulation and the gain of the grains also mapped to the Falcon's x and y coordinates. In this way, sound will only occur when the performer is hovering over a lighter area of the texture.

Additionally, all of the sound was sent through a noise gate. The threshold of the noise gate was determined by the z position of the Falcon, and simultaneously multiplied by the z force being sent to the Falcon. In this way, the fullest sounds can only be heard if the performer is hovering over a light, forceful area and if they are exerting a strong force on this area. Even if the performer applies the same force in a darker area, they will not be able to hear the fullest sound. It is only through a strong force in a physically demanding area that the performer can hear all of the frequencies at full volume. Even though there is a relatively large range between the weakest and strongest forces of the Falcon's resistance, it does not take a large amount of effort to merely slide over the different zones. The force-feedback is really only felt when the user pushes forwards. The Falcon is comparable to a joystick controller and, unlike a traditional instrument, its extremities are easily reached. According to Parker, ``grafting the joystick's physical extremes to the limits of software parameters can result in an unrewarding musical experience" \cite[p1]{parker15joys}. Therefore, it was more fruitful to explore the affordances of the physical extremities offered by the haptic terrains. Desirable sounds were still able to occur within the range of these limits. Multiplying the current resistance being output from the Falcon by the force received required the performer to exert significant effort in order to reach the full range of dynamics that the system had to offer. 

Isaac felt that this version of the sonic terrain was the most exciting to work with because of the diverse range of textures and resistances that it afforded. The interaction between gesture and sound seemed to correspond to the texture in a meaningful way. By mapping the combined resistance from the terrain and the performer's force to the threshold of the noise gate, dark areas of low resistance seemed to let small particles of the sound seep through, while light areas that required heavy force gave the sensation of physically pushing the full sound through the gate. The fact that feedback was no longer static, but became dependent on the response of the performer was also highly engaging.


\section{Conclusions and Future Work}
We have described the iterative development of a haptic DMI through the implementation of different strategies for generating force-feedback profiles. The first strategy involved a constant resistant force that made movement in one dimension more difficult. While this allowed the performer to more expressively explore regions of their sonic material in more detail, the uniformity of the resistance was deemed to be undesirable after numerous performances. The second strategy explored the idea of cross-modal terrains, generated as visual planes, and mapped to both haptic and sonic parameters. In particular, the Voronoi terrains generated by the basis functions afforded rich and varied haptic profiles.

These nonuniform terrains can be generated on the fly, allowing for instantaneous re-mapping to take place. Future work will explore this by changing force-feedback profiles during performance. The nodes-based model will also be repurposed as an audio processing routing strategy, running concurrently with the Voronoi technique. We will also experiment with projecting the terrains for the audience so that they can watch the performer attempt to navigate space both physically, as well as digitally. 



%
\bibliographystyle{abbrv}
\bibliography{nime-references}  

\begin{thebibliography}{10}

\bibitem{berdahl2009hsp}
E.~Berdahl, G.~Niemeyer, and J.~O. Smith.
\newblock {HSP}: A simple and effective open-source platform for implementing
  haptic musical instruments.
\newblock In {\em Proceedings of the International Conference on New
  Instruments for Musical Expression}, pages 262--263, 2009.

\bibitem{berdahl2007physically}
E.~Berdahl, B.~Verplank, J.~O. Smith~III, and G.~Niemeyer.
\newblock A physically-intuitive haptic drumstick.
\newblock In {\em Proceedings of the International Computer Music Conference},
  pages 363--366, 2007.

\bibitem{cadoz1988instrumental}
C.~Cadoz.
\newblock Instrumental gesture and musical composition.
\newblock In {\em ICMC 1988-International Computer Music Conference}, pages
  1--12, 1988.

\bibitem{filatriau2006using}
J.-J. Filatriau, D.~Arfib, J.-M. Couturier, and B.~Yeti.
\newblock Using visual textures for sonic textures production and control.
\newblock In {\em Proceedings of the International Conference on Digital Audio
  Effects}, pages 31--36, 2006.

\bibitem{hayes2011vibrotactile}
L.~Hayes.
\newblock Vibrotactile feedback-assisted performance.
\newblock In {\em Proceedings of the International Conference on New
  Instruments for Musical Expression}, pages 72--75, 2011.

\bibitem{hayes2012performing}
L.~Hayes.
\newblock Performing articulation and expression through a haptic interface.
\newblock In {\em Proceedings of the International Computer Music Conference},
  2012.

\bibitem{james2016multi-point}
S.~James.
\newblock Multi-point nonlinear spatial distributions of effects across the
  soundfield.
\newblock In {\em Proceedings of the International Computer Music Conference},
  2016.

\bibitem{leonard2015physical}
J.~Leonard and C.~Cadoz.
\newblock Physical modelling concepts for a collection of multisensory virtual
  musical instruments.
\newblock In {\em Proceedings of the International Conference on New
  Instruments for Musical Expression}, pages 150--155, 2015.

\bibitem{overholt2011advancements}
D.~Overholt, E.~Berdahl, and R.~Hamilton.
\newblock Advancements in actuated musical instruments.
\newblock {\em Organised Sound}, 16(02):154--165.

\bibitem{parker15joys}
M.~Parker.
\newblock Joys of travel: Introducing the spectral tourist.
\newblock {\em Leonardo Electronic Almanac}, 15.

\bibitem{tanaka2016haptic}
A.~Tanaka and A.~Parkinson.
\newblock Haptic wave: A cross-modal interface for visually impaired audio
  producers.
\newblock In {\em Proceedings of the ACM Conference on Human Factors in
  Computing Systems}, pages 2150--2161. ACM, 2016.

\end{thebibliography}
\end{document}